# Two-flavor staggered-fermion thermodynamics at $N_t = 12$ [*]

C. Bernard,[a] T. Blum,[b] C. DeTar,[c] Steven Gottlieb,[d] Urs M. Heller,[e] J. Hetrick,[b] K. Rummukainen,[d] R. Sugar,[f] D. Toussaint,[b] and M. Wingate[g]

[a]Department of Physics, Washington University, St. Louis, MO 63130, USA

[b]Department of Physics, University of Arizona, Tucson, AZ 85721, USA

[c]Physics Department, University of Utah, Salt Lake City, UT 84112, USA

[d]Department of Physics, Indiana University, Bloomington, IN 47405, USA

[e]SCRI, The Florida State University, Tallahassee, FL 32306, USA

[f]Department of Physics, University of California, Santa Barbara, CA 93106, USA

[g]Physics Department, University of Colorado, Boulder, CO 80309, USA

We present new results in an ongoing study of the nature of the high temperature crossover in QCD with two light fermion flavors. These results are obtained with the conventional staggered fermion action at the smallest lattice spacing to date—approximately 0.1 fm. Of particular interest are a study of the temperature of the crossover, an important indicator of continuum scaling, a determination of the induced baryon charge and baryon susceptibility, used to study the dissolution of hadrons at the crossover, the scalar susceptibility, a signal for the appearance of soft modes, and the chiral order parameter, used to test models of critical behavior.

## 1. INTRODUCTION

Lattice simulations of high temperature QCD have provided our only present firmly grounded theoretical insights into the phenomenology of the crossover from hadronic matter to the quark-gluon plasma and into the nature of the plasma. Outstanding problems include (1) establishing continuum scaling, important not only for determining the temperature of the crossover, but necessary for the validity of all dynamical fermion simulations, (2) establishing the mechanism for the dissolution of hadrons at the crossover, (3) exploring the intricate critical behavior associated with the phase transition, and (4) obtaining a quantitative characterization of the quark-gluon plasma, including the equation of state. To achieve these goals requires a combination of advances in algorithms and computing power [1]. Rapid improvements in lattice formulations may soon have a significant impact on dynamical fermion simulations [2]. Here we give a progress report on an analysis of simulations with the conventional staggered fermion action at the smallest lattice spacing to date, namely, approximately 0.1 fm [3].

The simulation with two flavors of quarks was carried out at two quark masses, $ma = 0.008$ and 0.016 and six couplings, $6/g^2 = 5.65$, 5.70, 5.725, 5.75, 5.80, 5.85, except that the 5.725 coupling was not simulated at the higher mass. In each case the simulation was extended to at least 2000 molecular dynamics time units. Lattices were saved at intervals of 8 time units. For the present analysis the first 500 time units were omitted. Most of the results reported here are based on an analysis of the approximately 180 remaining lattices at each parameter pair. Although results presented here are based on more data than reported at the conference, they should still be regarded as preliminary.

---

[*]Presented by C. DeTar

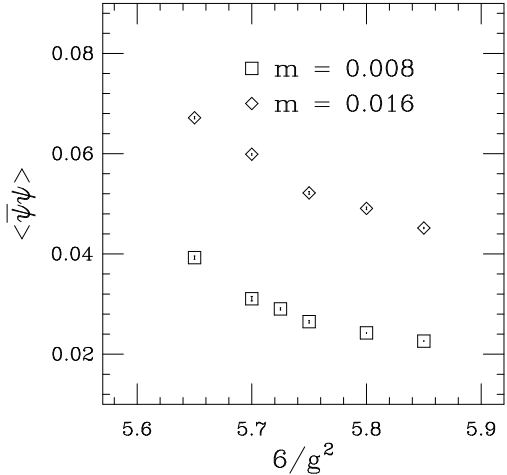

Figure 1. Chiral condensate $\langle\bar{\psi}\psi\rangle$ vs $6/g^2$.

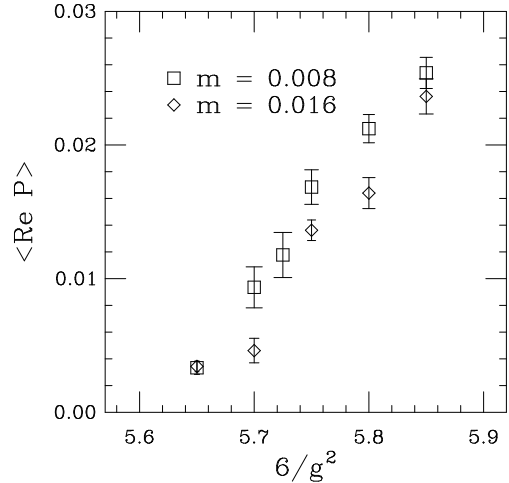

Figure 2. Polyakov loop vs $6/g^2$.

## 2. LOCATING THE CROSSOVER

### 2.1. Polyakov loop, $\langle\bar{\psi}\psi\rangle$, fuzzy loop

Plotted in Figs. 1 and 2 are the observables traditionally used to locate the crossover, which is signalled by an inflection point. Despite small errors, it is evident that locating an inflection point from these results is challenging, to say the least. To strengthen the crossover signal we tried constructing a "fuzzy" loop variable by replacing each time-like link $U_t$ in the conventional Polyakov loop with

$$U_{\text{fuzz},t} = \alpha U_t + \beta \sum U_{\text{staple}} \qquad (1)$$

where $U_{\text{staple}}$ is one of the six staples of length three links connecting the ends of the conventional single link [4]. This definition is not recursive and the resulting link is not projected onto SU(3). For present purposes we chose a weighting $\alpha = \beta = 1/7$. Although this variable is slightly less noisy than the Polyakov loop, the crossover signal remained unclear.

### 2.2. Baryon susceptibility

The conventional Feynman path integral simulates the grand canonical ensemble in baryon number at zero chemical potential. The baryon susceptibility measures fluctuations in the baryon number of the ensemble. It is defined as the derivative of the baryon charge density with respect to chemical potential. The susceptibility can be defined separately for each flavor. Thus with two quark flavors, two susceptibilities can be measured: a flavor singlet and flavor nonsinglet [5]. Both quantities are obtained at zero chemical potential.

$$\chi_{\text{s,ns}} = \left(\frac{\partial}{\partial\mu_u} \pm \frac{\partial}{\partial\mu_d}\right)(\rho_u \pm \rho_d) \qquad (2)$$

We present here only $\chi_{\text{ns}}$. Results are compared with those for lower $N_t$ in Fig. 3 [5]. Also indicated is the free lattice quark value for each $N_t$. A common pattern is the tendency for the susceptibility to rise abruptly at crossover and approach the free lattice quark value asymptotically. At lower values of $N_t$, where the crossover has been located with traditional methods, we find that the baryon susceptibility reaches 1/3–1/2 of the free quark asymptotic value at the crossover. Following the same rule for $N_t = 12$ places the crossover in the $ma = 0.008$ series between $6/g^2 = 5.65$ and 5.70 and in the $ma = 0.016$ series between $6/g^2 = 5.75$ and 5.80.

To convert these results to a temperature, we use a scale in which the rho meson mass is taken to be 770 MeV, regardless of quark mass. The rho mass in lattice units is obtained in turn from fitting a polynomial to masses in a compilation of two-flavor staggered fermion spectral simulations [6], and includes an extrapolation beyond the parameter range of spectral measurements ($6/g^2 > 5.7$) using tadpole-improved asymptotic scaling [3]. From the baryon susceptibility we then place the crossover at $T_c = 143 - 154$ MeV$_\rho$ at the lighter quark mass and $T_c = 142 - 150$ MeV$_\rho$ at the heavier quark mass. These values are consistent with results at smaller $N_t$ [3], so reveal no drastic departures from scaling.

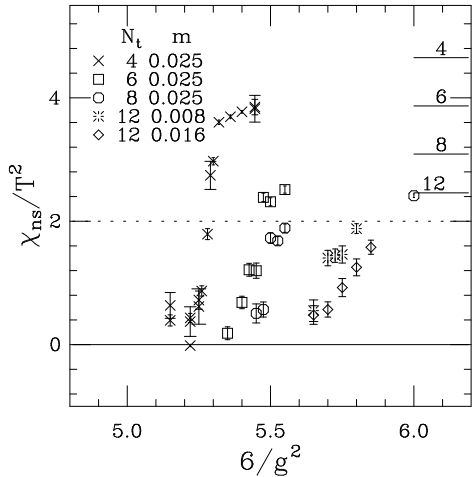

Figure 3. Nonsinglet baryon susceptibility *vs* $6/g^2$ for $N_t = 4, 6, 8, 12$. Free lattice quark values for each $N_t$ are indicated by horizontal lines on the right of the plot.

### 2.3. Induced quark number

Another confinement-sensitive observable is the induced quark number [7]. This observable measures the total residual light-quark number in an ensemble containing a single test quark.

$$Q_{\text{ind}} = \int \rho_{\text{ind}}(r) d^3 r \qquad (3)$$

The induced quark number density is measured in the presence of the test quark. Operationally, the quantity measures the correlation between the Polyakov loop and the light-quark density [7]. To reduce the considerable noise, we introduced the test charge through the fuzzy Polyakov loop variable described above. Results are shown in Fig. 4 and compared with results from simulations at lower $N_t$. The induced quark number is expected to be exactly $-1$ at zero temperature, since confinement requires screening of the test charge by a single antiquark. At the crossover, this quantity rises rapidly, approaching zero in the high temperature phase. At lower $N_t$ the induced quark number reaches approximately $-0.1$ at crossover. Applying this rule to the $N_t = 12$ data gives $6/g^2 = 5.65 - 5.70$ ($T = 143 - 154$ MeV$_\rho$) in the $ma = 0.008$ series and $6/g^2 = 5.70 - 5.80$ ($T = 134 - 150$ MeV$_\rho$) in the $ma = 0.016$ series. This crossover location is consistent, but somewhat less precise, than that found from the baryon susceptibility.

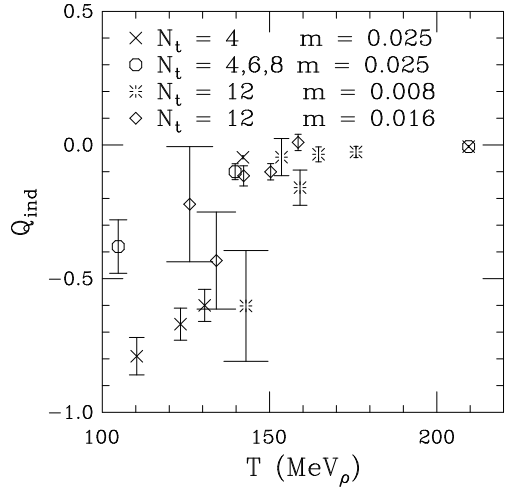

Figure 4. Induced quark number. Octagons are for fixed lattice spacing.

## 3. CHIRAL SUSCEPTIBILITY

Another signal for the crossover is the chiral susceptibility

$$\chi_m = \left. \frac{\partial \langle \bar{\psi}\psi \rangle}{\partial m} \right|_{6/g^2} \qquad (4)$$

which measures fluctuations in the chiral order parameter $\langle \bar{\psi}\psi \rangle$ [8]. Since it is the space-time integral of the correlator $\langle \bar{\psi}\psi(r)\bar{\psi}\psi(0) \rangle$, a peak in this observable occurs at a minimum in the $\sigma$ meson screening mass, indicating the presence of a soft mode. Such soft modes are expected in models of critical behavior [9]. Results for the quark-line disconnected contribution to this susceptibility are compared in Fig. 5 with the results of Karsch and Laermann at $N_t = 4$ [8]. The trend is consistent with a peak at the $6/g^2 = 5.70$ data point (153 MeV$_\rho$) in the lighter mass data and at the $6/g^2 = 5.80$ data point (150 MeV$_\rho$). However, the signal in the higher mass data is not strong enough to be predictive. That the signal should be weaker at higher quark mass is expected from models of critical behavior that place the critical point at zero quark mass. The mass ratio $m_\pi^2/m_\rho^2$ measures proximity to the critical point. In the $N_t = 4$ data, this ratio is less than 0.1 for the $ma = 0.020$ data and between 0.1 and 0.2 for the $ma = 0.0375$ data. For the $N_t = 12$ data, this ratio is between 0.2 and 0.3 for $ma = 0.008$ and 0.3 to 0.4 for $ma = 0.016$.

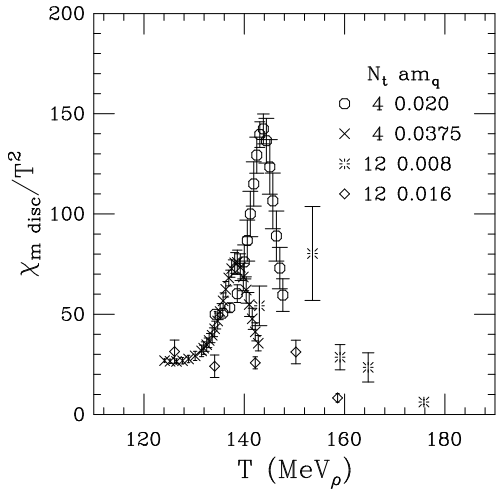

Figure 5. Disconnected chiral susceptibility for $N_t = 4$[8] and 12.

## 4. CONCLUSIONS

With the conventional staggered fermion action, locating the high temperature crossover at 0.1 fm for $m_\pi^2/m_\rho^2 > 0.2$ is difficult in the conventional observables $\langle \mathrm{Re}P \rangle$ and $\langle \bar{\psi}\psi \rangle$. However, the baryon susceptibility and induced quark number provide an adequate determination. The crossover temperature at the lighter quark mass is found to be in the range $143 - 154$ MeV$_\rho$, consistent with results at lower $N_t$. Other qualitative features of the crossover, seen at lower $N_t$, are confirmed. In particular, the crossover is signaled by an abrupt rise in baryon susceptibility and induced quark number.

This work was supported in part by the U.S. National Science Foundation and by the U.S. Department of Energy and was carried out at NCSA, SDSC, PSC, NERSC, and Sandia.